# Redefining the Phase Space for Ideal Gas Systems Resolves the Gibbs Paradox

Quanmin Guo

*School of Physics and Astronomy, University of Birmingham, Birmingham B15 2TT, UK*

**Abstract:** For an ideal gas consisting $N$ molecules within a volume $V$, the volume accessible to each molecule at an instantaneous time is $V/N$. The rest of the volume, $(N-1)(V/N)$, is occupied by other $(N-1)$ molecules. The textbook assumption that a molecule can access any location inside the volume $V$ at one instantaneous in time is wrong leading to the Gibbs paradox. By taking into account the correct physical space for individual molecules, the single molecule partition function is $Z_1 = \dfrac{V}{N\lambda^3}$. The partition function for the N-molecule system is simply $Z_N = Z_1^N = \left(\dfrac{V}{N}\right)^N \dfrac{1}{\lambda^{3N}}$ which gives rise to the correct entropy of the system. There is neither the need to introduce the $\dfrac{1}{N!}$ factor nor the requirement to argue about the distinguishability of molecules. Entropy of mixing two quantities of ideal gasses is zero no matter the gasses are the same type or different types. With the appropriate assignment of the phase space, the entropy of the system has the expected property of being extensive and the Gibbs paradox is removed.

## 1. Introduction

The entropy for an ideal gas consisting of $N$ molecules in a volume $V$ at temperature $T$, evaluated directly from the first law of thermodynamics, $dU = TdS - pdV$, has the familiar form:

$$S = \frac{3}{2}Nk\ln T + Nk\ln V + C, \qquad (1)$$

where $k$ is the Boltzmann constant. For closed systems where $N$ is fixed, $C$ is usually treated as a constant with little physical significance. However, as discussed by Jaynes [1] that the accurate form of entropy, when considering its dependence on the particle number, is:

$$S(T,V,N) = \frac{3}{2}Nk\ln T + Nk\ln V + kf(N), \qquad (2)$$

where $f(N)$ is a function of $N$. Applying the condition that entropy must be extensive, Pauli found a general solution of $f(N)$ [2]:

$$f(N) = Nf(1) - N\ln N, \qquad (3)$$

where $f(1)$ is a constant. Hence, the entropy expressed as a function of $T$, $V$ and $N$ is:

$$S(T,V,N) = Nk\left[\frac{3}{2}\ln T + \ln\frac{V}{N} + f(1)\right]. \qquad (4)$$

Equation (3) is assembled in order to make entropy an extensive quantity, rather than being derived from a physical ground. In this paper, we will demonstrate that Eq. (4) can be derived from first principles based on an appropriate assignment of the phase space to the molecules.

In the standard statistical mechanical treatment of canonical ensembles, the entropy of an ideal gas system in thermal equilibrium at temperature $T$ can be directly derived from the partition function $Z(T, V, N)$. For $N$ independent molecules, $Z(T,V,N) = Z_1^N$ where $Z_1$ is the single molecule partition function. However, the entropy derived from a such partition function has a well known problem of being non-extensive and it leads to the Gibbs paradox [3]. Many attempts have been made to rectify this problem and the popular solution found in many textbooks [4] is to use $Z(T,V,N) = Z_1^N \times \frac{1}{N!}$.

The introduction of the factor, $\frac{1}{N!}$, is claimed to be necessary for systems containing indistinguishable particles to remove over counted states. The legitimacy of the $\frac{1}{N!}$ factor together with the argument around distinguishability of particles has been subject to intensive debate [5-11]. Here, we show that the phase space assigned to a single molecule is incorrect in previous analysis leading to $Z_1$ which is too large. For a single molecule in a system of volume $V$ containing $N$ molecules, the physical space available at one instant in time is is $V/N$. This is based on the physical reality that the total volume $V$ is shared among the $N$ molecules. By taking into account the appropriate physical space belonging to each molecule, we would have $Z_1$ proportional to $V/N$. The partition function for the whole system is thus simply $Z_1^N$ which naturally leads to an extensive entropy without the need to introduce any correction factors.

## 2. Entropy of a classical system consisting of *N* identical particles

We first review the classical approach in finding the partition function for an *N*-molecule ideal gas system. The single molecule partition function is:

$$Z_1 = \frac{1}{h^3} \int d^3q \, d^3p \, e^{-\frac{p^2}{2mkT}}. \tag{5}$$

In the above equation, *q* is the position and *p* the momentum of the molecule, respectively. *h* is the Planck constant. The integral over position gives:

$$\int d^3q = V,$$

which is the volume of the box.

$$\int d^3p \, e^{-\frac{p^2}{2mkT}} = \int d^3p \, e^{-\frac{p_x^2+p_y^2+p_z^2}{2mkT}}$$
$$= \int dp_x e^{-\frac{p_x^2}{2mkT}} \int dp_y e^{-\frac{p_y^2}{2mkT}} \int dp_z e^{-\frac{p_z^2}{2mkT}} \tag{6}$$

The partition function for a single molecule is thus:

$$Z_1 = V \left( \frac{2\pi mkT}{h^2} \right)^{\frac{3}{2}} = \left( \frac{V}{\lambda^3} \right), \tag{7}$$

where $\lambda = \sqrt{\frac{h^2}{2\pi mkT}}$ is the thermal wavelength.

For *N* molecules within the same volume, we have:

$$Z_N = \frac{Z_1^N}{N!} = \frac{1}{N!} \frac{V^N}{\lambda^{3N}}. \tag{8}$$

The inclusion of $\frac{1}{N!}$ in the above equation is believed to be a necessary step to eliminate over counting of states. The over counting of states is illustrated in Figure 1 using the standard textbook argument for two molecules and two energy levels. Fig. 1(a) and (d) are clearly two different states. Fig. 1(b) and (c) show two other possible arrangements with one molecule on each level. According to textbook argument, (b) and (c) are the same because the two molecules are indistinguishable. No experimental procedure allows us to know which molecule occupies which level. Therefore, (b) and (c) are bundled together and treated as a single microstate. For the system containing two molecules, $Z_1^2$ gives four states, rather than three and hence correction is necessary to remove the redundant states. For a system with $N$ molecules, the factor $\frac{1}{N!}$ is introduced to account for the permutation of $N$ molecules.

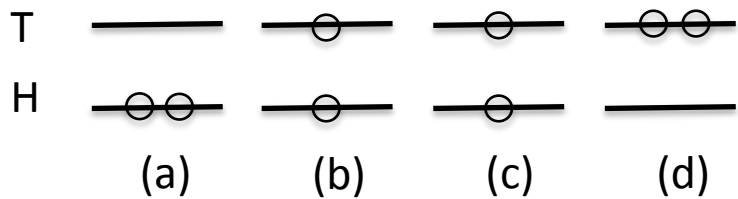

Figure 1. Counting the number of microstate using two particles as an example. We use H (heads) and T (tails) to represent two energy levels for each particle. There are a total of four possible states (a)-(d). States (b) has the same probability to occur as state (c). For two indistinguishable particles, state (b) and (c) cannot be distinguished experimentally and are thus counted as a single state. The combination of state (b) and (c) into a single state is flawed leading to incorrect statistics.

The argument of indistinguishability of particles and the hence the justification of the $\frac{1}{N!}$ factor is well accepted and appear in many textbooks. However, such an argument has a fundamental flaw and is subject to continued debate. Just because the two configurations in Fig. 1(b) and (c) are experimentally indistinguishable, we cannot conclude that they do not physically exist as two independent states. We can prove that (b) and (c) in Fig. 1 are two distinctive states by considering the flipping of two coins. We can have two identical coins that we are unable to tell which is which. The observed outcome from each throw is thus (H, H), (H, T), (T, H) or (T, T). We are unable to distinguish (H, T) from (T, H) because the two coins are exactly the same. However, (H, T) and (T, H) cannot be treated as one state. The probability that (H, T) occurs is ¼, so is the probability for (T, H). The probability that either (H, T) or (T, H) is observed is ½. If (T, H) and (H, T) are counted as a single state, the probability of observing (H, T) or (T, H) would be 1/3. State (b) and (c) in Figure 1 are thus two independent states, each carries its own statistical weight. If we were to consider these two states as a single state, we would have incorrect statistics. The introduction of $\frac{1}{N!}$ in Eq. (8) makes the system entropy an extensive quantity, but its introduction contradicts the principles of probability and statistical physics. Our analysis above demonstrates that the $\frac{1}{N!}$ should not be incorporated into Eq. (8). Without this factor, however, the partition function leads to a non-extensive entropy. The problem, as we will show in the following, is rooted in a wrong assumption made in the evaluation of the single molecule partition function $Z_1$. In the derivation leading to $Z_1$, Eq. (7), $\int d^3q = V$ is applied without justification. If there is only one molecule inside volume $V$, we have $\int d^3q = V$ because the molecule has equal probability to be found anywhere inside V. What happens if there are $N$ molecules in the same volume? Since we are dealing with non-interacting particles, the $N$ molecules are conventionally treated as independent from each other. It thus appears that each of the $N$ molecules can access any point inside $V$ independent of what happens to other molecules. However, the accessibility of a molecule to a particular point in $V$ depends on the existence of, and the number of, other molecules. At one instantaneous in time, each molecule has the freedom to access only a small fraction of the total volume $V$. On average, we expect each molecule has just $v = \frac{V}{N}$ of physical space. It is true that over a sufficiently long time, a molecule may have visited every single point in the physical space provided by $V$. But, at any one moment, the space that is available to a molecule is $v$ with the rest of the volume, $V - v$, occupied by other *(N-1)* molecules. The partition function does not include any integration over time. For any space that is already occupied by a molecule, that space is not accessible to other molecules at the same time. This kind of space exclusion must be true for all types of particles in statistical systems. The state of each particle is specified with $(x_i, y_i, z_i, p_{xi}, p_{yi}, p_{zi})$. There is no restriction on the momentum/energy of the particle, but no two particles can have the same $(x_i, y_i, z_i)$ coordinates. Therefore, we can rule out the possibility of two or more particles having the same state. The average volume per molecule, $v$, is rarely used in literature.

However, the molecular density, $n = \frac{1}{v} = \frac{N}{V}$, is a well-defined parameter and this fully justifies the use of $v$ in our discussion. At equilibrium, it is expected that for any arbitrary location in the system, the molecular density is constant subject only to fluctuation. Thus, one can make an equivalent statement that the average volume per molecule, $v$, is constant at any location in the system.

One can treat the whole system consisting of $N$ sub-systems, each sub-system has the same volume, $V/N$, and contains one molecule as shown in Figure 2. Fig. 2(a) represents a piece of solid where individual molecules are completely localized. Fig. 2(b) is for a system where each molecule is given some freedom to move around. Fig. 2(c) is close to the gaseous system where each molecule can move freely inside the whole volume $V$. During a short enough period of time, each molecule would have traveled inside a volume $V/N$ around a particular point. For instance, the mean free path of a molecule in atmosphere at room temperature is only around 60 nm. The average nearest neighbor distance of molecules in the air is ~ 10 nm. Thus, the physical space available to a single molecule in atmosphere is ~ $10^{-22}$ m$^3$. The partition function $Z_1$ for a molecule in such a system is thus:

$$Z_1 = \frac{1}{h^3} \int d^3q\, d^3p\, e^{-\frac{p^2}{2mkT}}$$
$$= \frac{V}{Nh^3} \int d^3p\, e^{-\frac{p^2}{2mkT}} = \left(\frac{V}{N\lambda^3}\right) \quad (9)$$

The partition function for $N$ particles follows directly from Eq (9):

$$Z_N = Z_1^N = \left[\frac{V^N}{N^N \lambda^{3N}}\right]. \quad (10)$$

From Eq. (10), we can get the internal energy,

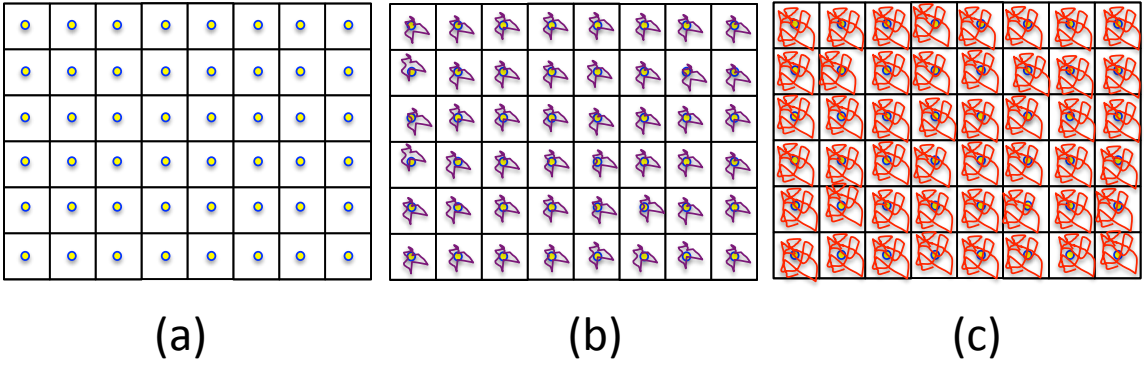

Figure 2. Illustration of an *N* particle system. (a) A piece of solid with *N* atoms with each atom completely localized. (b) Each atom has some limited freedom to move around its equilibrium position. (c) A "gas". During a very short time duration, each atom/molecule is perceived as free moving inside a volume *V/N*.

$$U = \frac{3}{2}NkT, \quad (11)$$

the Helmholtz free energy,

$$F = NkT \ln\left(\frac{N}{V}\lambda^3\right), \quad (12)$$

and the entropy,

$$\begin{aligned}S &= Nk\left[\frac{3}{2} + \ln\left(\frac{V}{N\lambda^3}\right)\right] \\ &= Nk\left[\frac{3}{2} + \ln\left(\frac{V}{N}\right) + \frac{3}{2}\ln\left(\frac{2\pi mkT}{h^2}\right)\right]\end{aligned} \quad (13)$$

Eq. (13) above has the same form as Eq. (4) given by Pauli with $f(1) = \frac{3}{2}\left(1 + \ln\frac{2\pi mk}{h^2}\right)$. Here, the equation is derived by considering the division of the physical space by *N* molecules. The entropy of the system given by Eq. (13) is extensive.

Note, we are dealing with *N* independent molecules confined within volume *V*. The single molecule partition function $Z_1$ and the N-molecule partition function $Z_N$ is simply related by $Z_N = Z_1^N$, as long as the single molecule partition function is correctly evaluated. There is no need to introduce any correcting factor. To extend the above discussion, lets consider a system which consists of *N* boxes each of volume *V*. There is one molecule inside each box and the whole system is under thermal equilibrium. The partition function for each box should be given by Eq. 7 as $Z_1 = V\left(\frac{2\pi mkT}{h^2}\right)^{\frac{3}{2}} = \left(\frac{V}{\lambda^3}\right)$. This leads to the partition function of the total system as $Z_N = Z_1^N = \frac{V^N}{\lambda^{3N}}$. One can see that this is also the partition function for *N* molecules in a volume of *NV*.

When $\frac{1}{N!}$ is introduced to $Z_N$, an assumption is usually made that no two molecules can have the same state. Otherwise, the correction factor is not sufficient. Explanation is not always given in textbooks why molecules having the same state can be ignored. It is sometimes argued qualitatively that the number of states well exceeds the number of particles, hence the probability that two particles having the same state is negligibly small. We have demonstrated that two molecules are strictly not allowed to have the same state because they must have different $(x_i, y_i, z_i)$ coordinates. This is a fundamental physics requirement that two molecules should not be overlapping in physical space, and this does not depend at all on the number of states. Even if there are just two molecules in the same volume, they are not allowed to have the same state. When Eq. (8) is used to calculate the entropy, Stirling's approximation is applied to $\ln N!$. This generates a question regarding the validity of such a procedure for small systems. In our analysis, no Stirling's approximation is required and thus Eq. (13) is expected to be valid for both large and small systems.

It is noted that in the standard textbook approach to solid state materials, the partition function is written as $Z_N = Z_1^N$ without the $\frac{1}{N!}$ correcting factor. For example, in the Einstein model of solids, $Z_{3N} = Z_1^{3N}$ for *3N* oscillators. It is usually argued that atoms in a solid are distinguishable because of their distinctive (*x,y,z*) coordinates and thus there is no over counting of states. The truth is, for a solid, we always assume that the volume is constant. Thus, the volume dependent part of the entropy is treated as a constant or just zero for convenience. In this case, we are satisfied that each atom is localized and is unable to move away from its fixed location.

## 3. Entropy of mixing

Entropy of mixing is one of the debated issues around the Gibbs paradox. Consider the typical system shown in Figure 3. The partition divides the total volume of *2V* into two halves. If each half contains *N* identical molecules at the same temperature *T*, when the partition is removed, entropy change according to Eq. (13) is zero. For the same reason, if the partition is inserted back there is no reduction in entropy leading to no paradox.

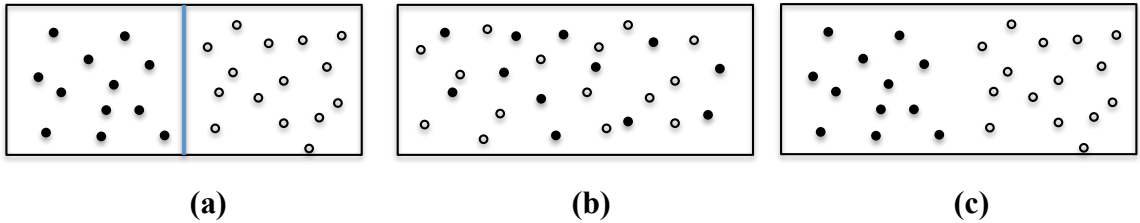

(a)  (b)  (c)

Figure 3. (a) A box with a volume *2V* is divided by a partition into two equals halves. Each half contains *N* particles. (b) Complete mixing of the particles after the partition is removed. The diagram shows a snap shot of the system and hence can be considered as one of the possible microstate of the system. (c) Another possible microstate of the system.

We can also analyze the situation where the molecules on the two sides are different, say molecule A on the left with mass $m_A$ and molecule B on the right with mass $m_B$. The two sides are initially under thermal and mechanical equilibrium. The density of molecules on each side is the same. The chemical potentials for the two sides are:

$$\mu_A = -kT \ln\left[\left(\frac{2\pi m_A kT}{h^2}\right)^{\frac{3}{2}} \frac{V}{N}\right] \text{ and } \mu_B = -kT \ln\left[\left(\frac{2\pi m_B kT}{h^2}\right)^{\frac{3}{2}} \frac{V}{N}\right],$$ respectively. If the partition is removed, mixing of the two molecules will lead to new equilibrium when the molecules A and B are uniformly distributed over the whole 2 V volume. The change in entropy due to this mixing is given by:

$$\Delta S = Nk\left[\frac{3}{2} + \ln\left(\frac{V_A}{N}\right) + \frac{3}{2}\ln\left(\frac{2\pi m_A kT}{h^2}\right)\right] + Nk\left[\frac{3}{2} + \ln\left(\frac{V_B}{N}\right) + \frac{3}{2}\ln\left(\frac{2\pi m_B kT}{h^2}\right)\right]$$

$$-Nk\left[\frac{3}{2} + \ln\left(\frac{V}{N}\right) + \frac{3}{2}\ln\left(\frac{2\pi m_A kT}{h^2}\right)\right] - Nk\left[\frac{3}{2} + \ln\left(\frac{V}{N}\right) + \frac{3}{2}\ln\left(\frac{2\pi m_B kT}{h^2}\right)\right]$$

$$= Nk\left[\ln\left(\frac{V_A}{N}\right) + \ln\left(\frac{V_B}{N}\right) - 2\ln\left(\frac{V}{N}\right)\right]$$

$$= Nk \ln\left(\frac{V_A}{N}\frac{V_B}{N}\left(\frac{N}{V}\right)^2\right) = Nk \ln \frac{V_A V_B}{V^2} \tag{14}$$

In the above equation, $V_A$ and $V_B$ are the volume "owned" by molecule A and B, respectively. Treating A and B as ideal gasses, it is expected that $V_A = V_B = V$. The total volume of 2$V$ is equally shared between the two kinds of molecules with each molecule having an equal share of the volume. Therefore, we have $\Delta S = 0$, i. e. the entropy of mixing two different gasses is zero. The conclusion we can draw from the above analysis is that when ideal gasses at the same temperature and pressure are mixed, the entropy of the combined systems remain constant. This applies to the mixing of two quantities of different ideal gasses or the mixing of two quantities of the same gas. Our result may seem surprising as many would expect a non-zero change in entropy when two dissimilar gasses are mixed. Here, we explain why the entropy remains constant. From a basic thermodynamics point of view, during mixing, there is no heat exchange with the surrounding. There is no change to the internal energy, so temperature remains constant. Hence, $dS = \frac{\delta Q}{T} = 0$ at anytime during mixing because $\delta Q = 0$. Mixing can be conducted in a quasi-static manner to ensure the process is reversible. The pressure, temperature and total volume of the system remains the same during mixing. Thus the system is under the same macrostate and entropy remains constant because it is a function of state. We will discuss the issues relating to the change of microstates later and explain why mixing is sometimes viewed as a spontaneous irreversible process based on the evolution of microstates.

The mixing of the two gasses is frequently described as a combination of free expansions of A from $V$ to $2V$ and B from $V$ to $2V$ yielding an overall entropy increase of

$$\Delta S = Nk \ln\left(\frac{2V}{V}\right) + Nk \ln\left(\frac{2V}{V}\right) = 2Nk \ln 2. \qquad (15)$$

This sequence of expansions has been incorrectly described as equivalent to diffusive mixing. We will show in the following that the entropy change from Eq. (15) is due to isothermal expansion, not due to mixing of the gasses. Figure 4(a) shows two equal volumes of ideal gasses A and B at the same temperature. By allowing each gas to expand from $V$ to $2V$, state (b) is created. Changing from (a) to (b) leads to an entropy increase which is given by Eq. (15). Although this entropy change has been assigned to the entropy of mixing in literature, we can see from Figure 4 that mixing has not occurred. We can take an extra step to mix the gasses. We make a small hole in the wall separating A and B as shown in Fig. 4(c), and slowly compress B using the piston at the right. The compression is conducted slowly enough such that the pressure on the two sides of the wall remains the same at all times. When all the B molecules from the right hand side have moved to the left, complete mixing is completed and state (d) is reached. From (c) to (d) there is a reduction in the entropy of the system by:

$$\Delta S = Nk \ln\left(\frac{2V}{V}\right) + Nk \ln\left(\frac{2V}{V}\right) = 2Nk \ln 2$$

This is obtained simply by considering the compression of an ideal gas from $4V$ to $2V$. Thus, the whole process from (a)-(b)-(c) to (d) causes zero change in entropy. State (d) can be reached directly from (a) by removing the partition wall and allow molecules to diffuse. Therefore, we can conclude that mixing along the path (a)-(d) results in zero change of entropy.

The above analysis demonstrates that treating the mixing process as two separate isothermal expansion steps is not correct. It is noted that in the isothermal expansion processes, Fig. 4(a)-(b), the pressure of the gasses decreases as the volume increases. In a mixing process, A molecules diffuse to the right and B molecules diffuse to the left. There is no change to the pressure of the system at any stage of mixing.

The mixing is a process of positional exchange inside the total volume and no work is involved. The mixing of two ideal gasses is in principle the same as the mixing of two quantities of the same ideal gas.

From the statistical point of view, it looks that mixing has increased the number of accessible states. By removing the partition, the accessible physical space for each molecule seems to have doubled. However, as we have demonstrated earlier, the true accessible space for each molecule is $V/N$. Changing from (a) to (b) as shown in Fig. 3 does not change the average accessible physical space for any molecule in the system. At equilibrium, the system moves from one microstate to another without changing its macrostate. Due to the very large number of microstates under the umbrella of a single macrostate, the system is unlikely to return to a particular microstate which it has already gone through. For this reason, there is a sense of irreversibility in mixing. For example, removing the partition in Fig. 3(a), the system is expected to spontaneously change into (b). This process seems to be irreversible because one does not believe that the system would move from (b) to (a). In fact, Figs. 3(b) and(c) can be viewed as two snapshots of the system in two different microstates. According to statistical physics, these two microstates have the same probability to appear. It is wrong to think that the state (c) has a lower probability to occur than (b). Because of the very large number of accessible microstates, the system is constantly on the move from one microstate to another, and such a move is microscopically irreversible, but macroscopically reversible.

For real gasses, the intermolecular potential energy depends on the properties of the molecules. As a result, when two real gasses mix, there involves energy exchange such that the system may release or absorb energy from the surrounding. In this case, the entropy of mixing is non-zero. All gasses are real gasses, thus, mixing always causes a non-zero entropy change unless the gasses are ideal. How much entropy change comes from mixing depends on the specific combination of the two gasses involved. For two gasses with extremely similar physical properties, the entropy of mixing is low.

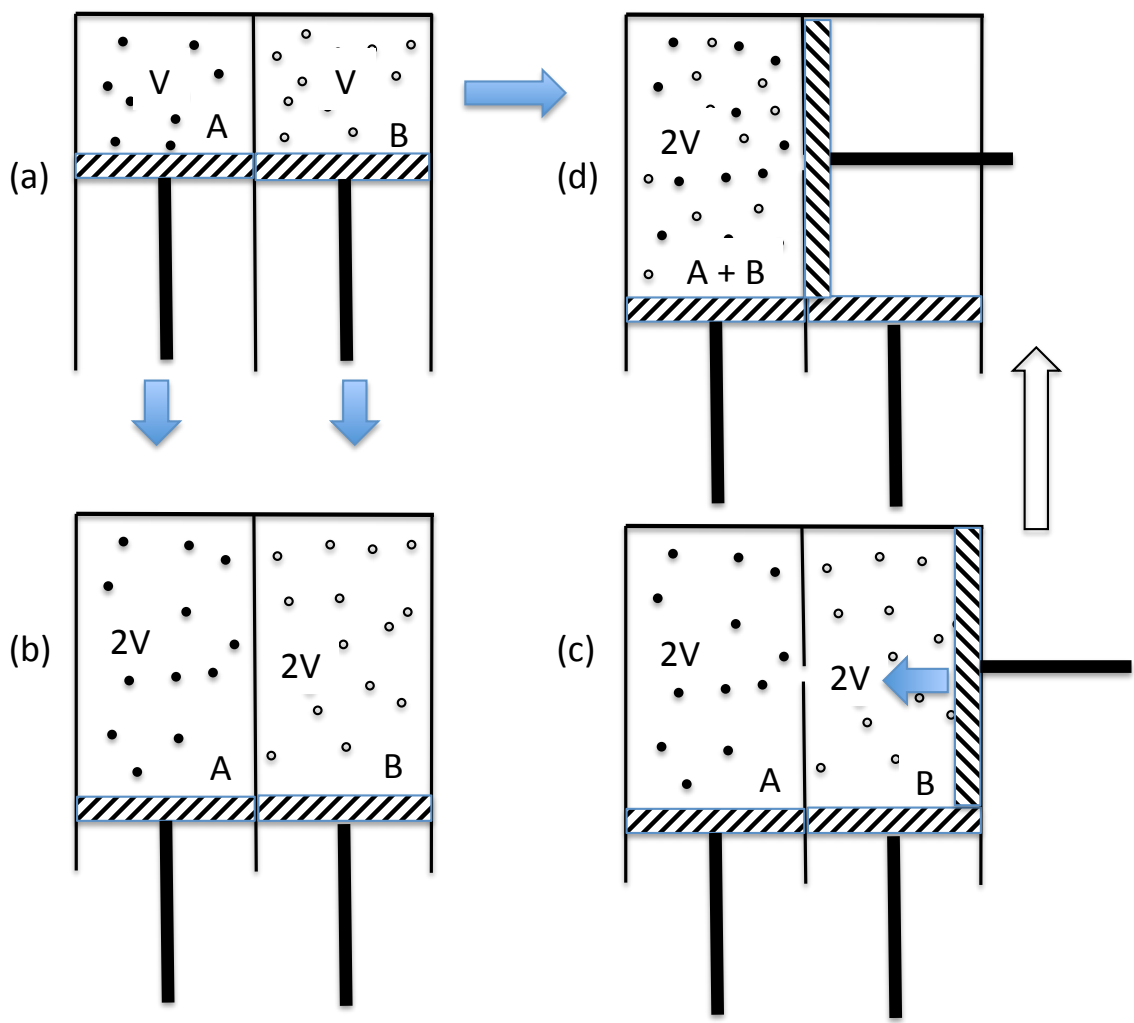

Figure 4. Mixing of two ideal gasses. (a) Gas A and B separated by a partition wall. (b) Each gas has expanded from V to 2 V. The two gasses are still under equilibrium. (c) Mixing the two gasses via isothermal compression by a piston. A small opening is made in the partition wall to allow molecules to pass through. (d) The two gasses in the final mixed state. Mixing can be conducted from (a) to (d) by removing the partition.

## 4. Discussion

The fundamentals of statistical physics treat all individual particles as indistinguishable. In analyzing a thermal system with a large number of particles, there is no need to know which particle is where and has what energy. Identification of individual particles is unnecessary. Even if each particle has a specific feature to allow us to recognize all individual particles, we would choose to ignore this information. All that needed are the number of particles, the volume and temperature of the system. Whither the particles are distinguishable or not does not affect the statistical understanding of the system. This is exactly the reason that we can apply statistical physics equally to atoms, molecules as well as the relatively large colloidal particles [6]. Our ability to separate one colloidal particle from another, and inability to tell one atom from another do not present a problem in analyzing the system using the same piece of physics. In the case of distinguishable colloidal particles, if we assume that each particle has full access to the whole physical volume of the system, we would encounter the same problem of a non-extensive entropy. Once we applied the condition that each particle has only $V/N$ space available, we end up with consistent outcomes for all particles, small or large, distinguishable or not. The available physical space for each particle is $V/N$ instead of $V$ is manifested even more clearly with the colloidal systems. Two colloidal particles cannot occupy the same $(x,y,z)$ coordinates in space and hence no two particles can have the same state. It is also not hard to imagine that from the total of volume $V$, only $V/N$ belongs to a single particle on average.

Accepting that each particle in the system is allocated a volume of $V/N$, all the thermodynamic quantities can be derived correctly. More importantly, the entropy of the system naturally comes out as an extensive quantity. The entropy of mixing becomes zero for ideal gasses no mater mixing is between two different gasses or between two quantities of the same gas. The Gibbs paradox is completely removed.

## 5. Conclusion

The physical volume accessible for a molecule in an ideal gas is $V/N$. The volume, $V/N$, for a gas molecule is not permanently localized. Over time, the volume associated with each molecule drifts around within the total volume $V$ of the system. At any instantaneous in time, the average physical space available to a molecule is $V/N$. The assumption that every single molecule has freedom to access any physical space provided by the total volume $V$ of the system is incorrect and it leads to over counting the number of states by including physically non-accessible states and hence the Gibbs paradox. Each molecule in the system has its own physical space and sharing this space with another molecule is not allowed. This is a direct consequence that molecules tend to spread over the given physical volume. Using the proper phase space for a molecule leads to no paradoxes and an expression of entropy that applies equally to distinguishable and indistinguishable particles. There is in fact no need to discuss distinguishability anymore because statistical physics intrinsically treats all particles as indistinguishable. If the trajectories of individual particles could be followed all the time, these trajectories would not offer anything extra to the analysis of the system except providing evidence supporting what is already written in the script of statistical and thermal physics.